\documentclass[seceq]{ptptex}
\usepackage{wrapft}
\usepackage{epsfig,multicol}

\newcommand\fverb{\setbox\pippobox=\hbox\bgroup\verb}
\newcommand\fverbdo{\egroup\medskip\noindent%
			\fbox{\unhbox\pippobox}\ }
\newcommand\fverbit{\egroup\item[\fbox{\unhbox\pippobox}]}

\newcommand {\beq}{\begin{equation}}
\newcommand {\eeq}{\end{equation}}
\newcommand {\beqa}{\begin{eqnarray}}
\newcommand {\eeqa}{\end{eqnarray}}

\notypesetlogo                       %comment in if to eliminate PTPTeX 
\title{%        %You can use \\ for explicit line-break
Planar Dominance in Non-commutative Field Theories\\ 
at Infinite External Momentum
}

\renewcommand{\thefootnote}{\fnsymbol{footnote}}

\author{%       %Use \scshape  for the family name
Tadayuki \textsc{Konagaya}$^{1,}$
\footnote{E-mail : konagaya@post.kek.jp}
and Jun \textsc{Nishimura}$^{1,2,}$
\footnote{E-mail : jnishi@post.kek.jp} 
}

\inst{%     %Affiliation, neglected when [addenda] or [errata]
$^1$Department of Particle and Nuclear Physics,\\
The Graduate University for Advanced Studies (Sokendai),\\
Tsukuba 305-0801, Japan\\
$^2$High Energy Accelerator Research Organization (KEK),\\
Tsukuba 305-0801, Japan
}

\abst{%         %this abstract is neglected when [addenda] or [errata]
In the perturbative expansion of field theories on a 
non-commutative geometry, it is known that planar diagrams dominate when
the non-commutativity parameter $\theta$ goes to infinity.
We investigate whether this ``planar dominance'' occurs also in the case that
$\theta$ is finite, but the external momentum goes to infinity instead.
While this holds trivially at the one-loop level, it is not obvious
at the two-loop level, in particular in the presence of UV divergences.
We perform explicit two-loop calculations
in the six-dimensional $\phi^3$ theory and confirm
that nonplanar diagrams after renormalization do vanish in this limit.
}

\begin{document}

\maketitle

\setcounter{footnote}{0}
\renewcommand{\thefootnote}{\arabic{footnote}}

\section{Introduction}

Non-commutative geometry \cite{Sny}
has been studied for quite a long time
as a simple modification of our notion of space-time
at small distances, possibly due to effects of 
quantum gravity \cite{gravity}.
It has recently attracted much attention, since it was shown that Yang-Mills theories
on a non-commutative geometry appear as the low energy limit of string theories with some background 
tensor field \cite{rf:SW}.
At the classical level, introducing non-commutativity to
the space-time coordinates modifies the ultraviolet dynamics
of field theories, but not the infrared properties.
This is not the case at the quantum level, however, 
due to the so-called UV/IR mixing effect \cite{rf:MRS}.
This effect causes various peculiar
long-distance phenomena, such as the spontaneous breaking of translational
invariance. In the scalar field theory, this phenomenon is
predicted in Ref.~\citen{GuSo} and confirmed by Monte Carlo
simulations in Refs.~\citen{Bietenholz:2002vj,AC,Bietenholz:2004xs}.
An analogous phenomenon is also predicted in gauge theories 
\cite{NCgauge}.
%The relation between the $\theta \rightarrow \infty$ limit 
%of noncommutative
%field theories and the large $N$ field theories are reconsidered
%in ref.~\citen{Bietenholz:2004as}.

In this paper, we focus on the
ultraviolet properties of non-commutative field theories.
In the perturbative expansion of field theories on a 
non-commutative geometry, planar diagrams dominate when
the non-commutativity parameter $\theta$ goes to infinity \cite{rf:MRS}.
This may be regarded as
a manifestation of the {\em nonperturbative} relation 
between the $\theta \rightarrow \infty$ limit 
of non-commutative field theories 
and the large $N$ matrix field theories \cite{Bietenholz:2004as},
which is based on the lattice formulation of 
non-commutative field theories \cite{AMNS}
and the Eguchi-Kawai equivalence 
\cite{Eguchi:1982nm,Gonzalez-Arroyo:1982hz}.
%(Note, however, that 
%this relation does not hold 
%when the translational invariance is spontaneously broken.)
We investigate whether ``planar dominance'' occurs also in the case that
$\theta$ is finite, but the external momentum goes to infinity instead.
While this holds trivially at the one-loop level \cite{rf:MRS}, 
it is not obvious at the two-loop level, in particular in the presence 
of UV divergences. We perform explicit two-loop calculations
in the six-dimensional $\phi^3$ theory and confirm
that nonplanar diagrams after renormalization do vanish in this limit.
We consider the massive case specifically, because in the massless
case, the equivalence of the infinite momentum limit and the
$\theta \rightarrow \infty$ limit follows from dimensional arguments.

Some comments on related works are in order.
In Ref.~\citen{Ishibashi:1999hs}, correlation functions of
Wilson loops in non-commutative
gauge theories are studied, and it is found
that planar diagrams dominate when the external momenta become large.
However, this result is based on a regularized theory, and the
issue of removing the regularization has not been discussed.
In Ref.~\citen{rf:BHN}, Monte Carlo simulations of a 2d non-commutative 
gauge theory are studied, and the existence of a sensible continuum 
limit is confirmed.
%nonperturbative renormalizability
There, the result for the expectation value of the Wilson loop agrees with the result of 
large $N$ gauge theory for small area, which implies the planar dominance
in the ultraviolet regime.
The aim of the present work is to confirm the planer dominance by explicit
diagrammatic calculations in a simple model taking account of
possible subtleties that arise at the two-loop level.
%due to planar subdiagrams in nonplanar diagrams.
Two-loop calculations in scalar field theories
are performed also in Refs.~\citen{rf:phi4}
and \citen{rf:phi3} in the case of $\phi ^4$
and $\phi ^3$ interactions, respectively, with different motivations.
The issue of renormalizability to all orders in perturbation theory
is discussed in Ref.~\citen{Chepelev:1999tt}.

The rest of this paper is organized as follows.
In \S \ref{section:renorm}
we define some notation necessary for the perturbative expansion
in non-commutative $\phi ^3$ theory.  
In \S \ref{section:type1} and \S \ref{section:type2}
we investigate nonplanar two-loop diagrams of different types separately
and show that the diagrams vanish in the $p^2 \to \infty $ limit.
%in which the an external line and an internal line cross.
%% We show that the diagram vanishes in the $p^2 \to \infty $ limit
%% after renormalizing the divergence.
%% In section 
%% we calculate the nonplanar diagram in which the internal lines cross.
%%  ,and we observe that this diagram
%% vanishes in the limit $p^2 \to \infty $ .
Section \ref{section:summary} is devoted to a summary and discussion.

\section{Perturbative expansion in non-commutative $\phi^3$ theory}
\label{section:renorm}

The Lagrangian density for the non-commutative $\phi^3$ theory 
in $d$-dimensional Euclidean space-time can be written as
\beq 
 \mathcal{L} = \frac{1}{2}(\partial_{\mu}\phi)^2 + \frac{m^2_{0}}{2}\phi^2
 + \frac{g_{0}}{3}\phi \star \phi \star \phi \ .
\eeq
Here, the $\star$-product is defined by
\beq
\phi (x) \star \psi (x) = \phi (x) \exp \Big( \frac{i}{2}
\overleftarrow{\partial_{\mu}} \theta_{\mu \nu} 
\overrightarrow{\partial_{\nu}} \Big) \psi (x) \ ,
%
%% \left. \bigl ( \phi \star \psi \bigr)(x)  
%% = \exp\Bigl[\frac{i}{2} \theta^{\mu \nu} 
%% \frac{\partial}{\partial x_{\mu}} 
%% \frac{\partial}{\partial y_{\nu}}  \Bigr]
%% \phi(x) \psi(y)\right|_{x=y} \ ,
\eeq
where $\theta ^{\mu \nu }$ is an antisymmetric tensor,
which characterizes the non-commutativity of the space-time.
The parameters $m_{0}$ and $g_{0}$ are 
the bare mass and the bare coupling constant, respectively.
As in the standard perturbation theory,
we decompose the bare Lagrangian density into the
renormalized Lagrangian density $\mathcal{L}_{\rm r}$
and the counterterms $\mathcal{L}_{\rm ct}$ as
$\mathcal{L}  =  \mathcal{L}_{\rm r} + \mathcal{L}_{\rm ct}$,
where
\beqa
\mathcal{L}_{\rm r}
&=& \frac{1}{2}(\partial_{\mu}\phi_{r})^2 + 
\frac{m^2}{2}\phi^2_{r} + \frac{g}{3}\phi_{r} \star \phi_{r}
\star \phi_{r}  \ , \\
\mathcal{L}_{\rm ct} &=&
\frac{1}{2}\delta_{z}
(\partial_{\mu}\phi_{r})^2 + 
\frac{\delta_{m}}{2}\phi^2_{r} + \frac{\delta_{g}}{3}
\phi_{r} \star \phi_{r} \star \phi_{r} \ .
\eeqa
Here we have introduced the following notation:
\beq
\phi \equiv  Z^{\frac{1}{2}}\phi_{r} \ , \quad \quad 
Z \equiv  1 + \delta_{z} \ ,  \quad \quad
\delta_{m} \equiv  m_{0}^2 Z - m^2 \ , \quad \quad 
\delta_{g} \equiv  g_{0}Z^{\frac{3}{2}}-g \ .
\eeq

\begin{figure}[htb]
\begin{center}
\includegraphics[width = 10cm,clip]{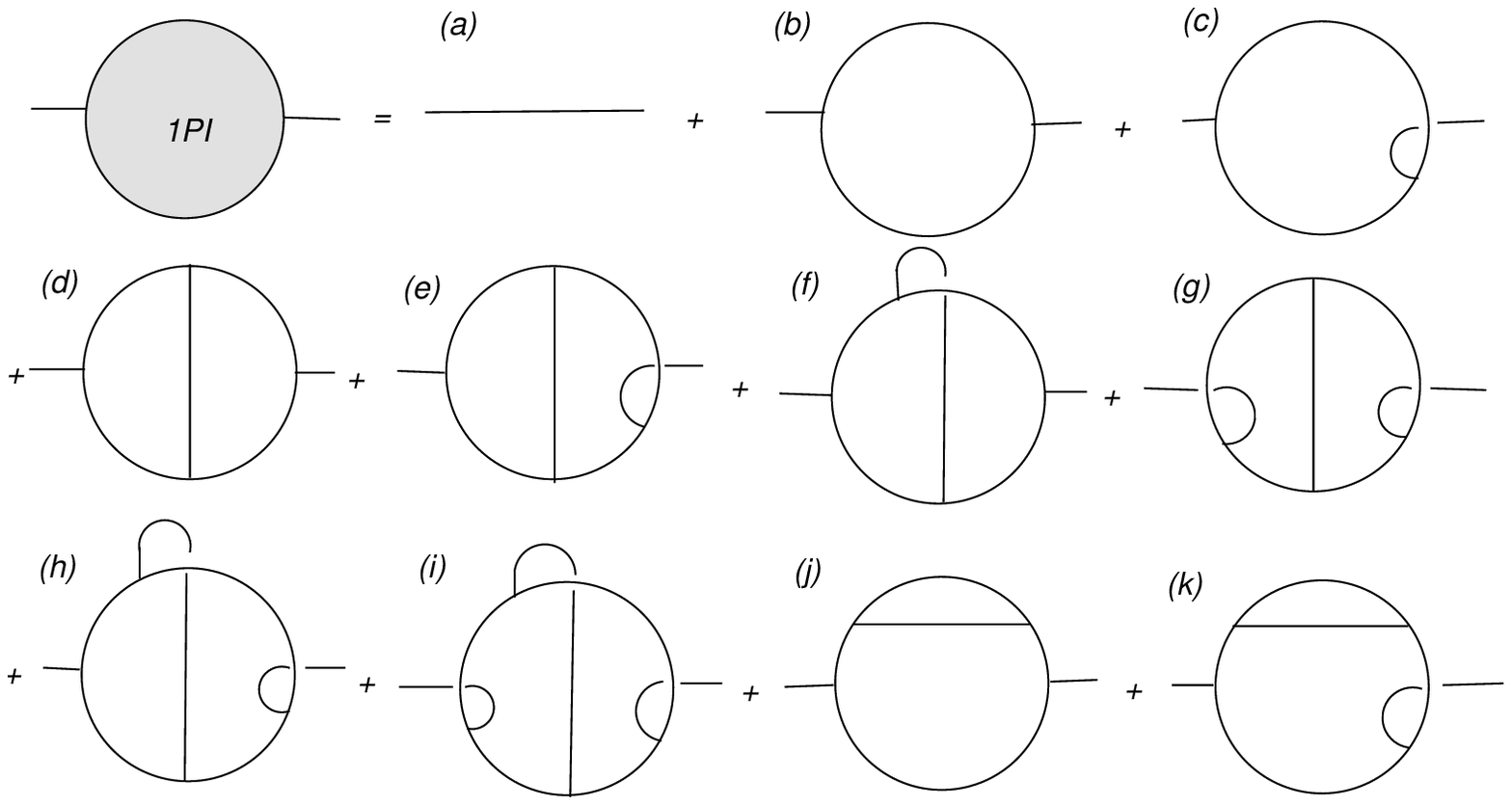}
\includegraphics[width = 10cm,clip]{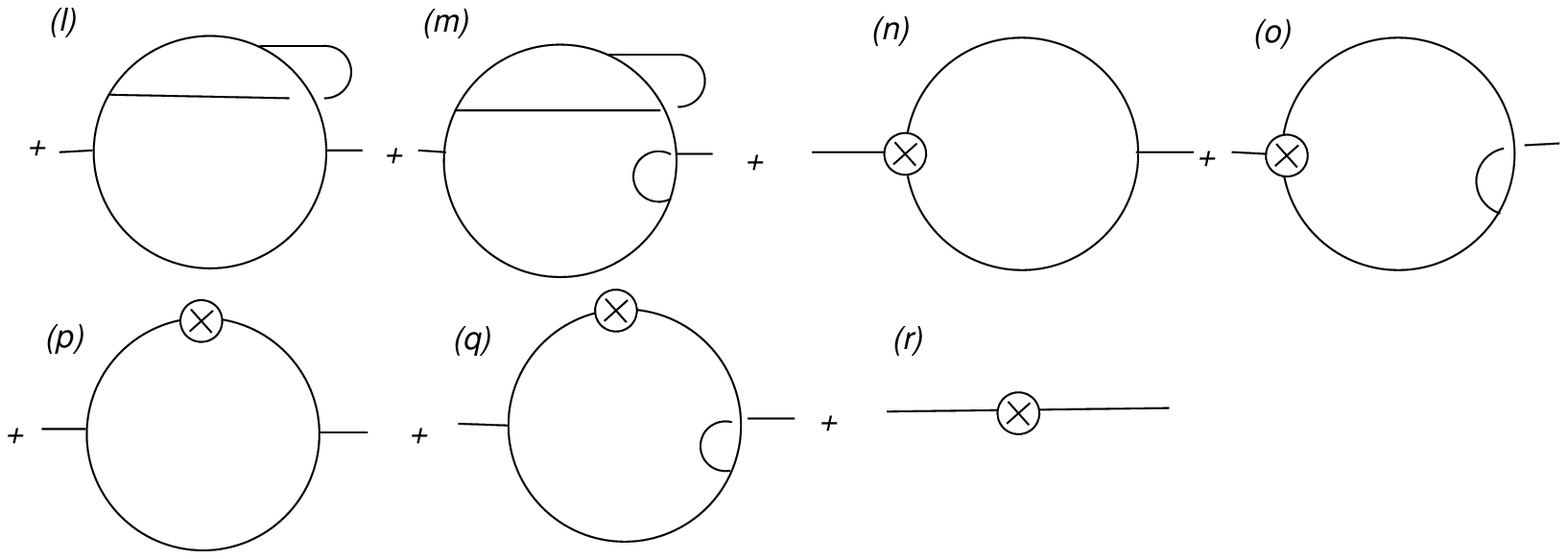}
\end{center}
\caption{List of diagrams for calculating the two-point 
function up to the two-loop level.}
\label{fig:two-point}
\end{figure}

The diagrams that need to be evaluated
in the two-loop calculation of the two-point function
are listed in Fig.\ \ref{fig:two-point}.
( We refer the reader to Ref.\ \citen{rf:MRS} for the Feynman rules.)
For diagrams (e),(h),(j),(k),(l),(m),(n),(o),(p) and (q),
we need to include a factor of 2 to take into account 
 the same contributions from analogous diagrams.
%to the same contributions from analogous diagrams.

%The number in the parenthesis assigned to each diagram represents
%a multiplicative factor due
%to the same contributions from analogous diagrams.
%
%%\begin{figure}[htb]
%%\begin{center}
%%\includegraphics[width = 10cm,clip]{twp1.eps}
%%\includegraphics[width = 10cm,clip]{twp2.eps}
%%\end{center}
%%\caption{jkljd;asldfj}
%%\label{twp1}
%%\end{figure}

%% \begin{figure}[htb]
%% \begin{center}
%% \includegraphics[width = 10cm]{nonplanarn.eps}
%% \end{center}
%% \caption{The two types of two-loop diagram studied in this paper.}
%% \label{nonplanarn}
%% \end{figure}

We focus on two types of nonplanar diagrams,
diagrams (e) and (f) in Fig.\ \ref{fig:two-point}.
%depicted in Fig.\ \ref{nonplanarn}.
The type 1 diagram is
a diagram in which an external line 
crosses an internal line.
This type includes the ultraviolet divergence coming
from the planar one-loop subdiagram.
We investigate whether this diagram vanishes 
at infinite external momentum
after appropriate renormalization.
The type 2 diagram 
is a diagram in which internal lines cross.
Since the non-commutativity parameter $\theta_{\mu\nu}$
does not couple directly to the external momentum in this case,
it is not obvious whether the effect of sending the external momentum
to infinity is the same as that of sending $\theta_{\mu\nu}$ to infinity.
%We will therefore study a representative two-loop diagram for each type
%depicted in Fig.\ \ref{nonplanarn}.
We comment on the remaining non-planar diagrams in 
\S \ref{section:summary}.

Throughout this paper, we consider the massive case
specifically. Technically, this condition 
simplifies the evaluation of the upper bound on the nonplanar diagrams.
Theoretically, this is the more nontrivial case,
because in the massless case,
the equivalence of the infinite momentum limit and the
$\theta \rightarrow \infty$ limit follows from dimensional arguments.

\section{Type 1 diagram}
\label{section:type1}

Because the type 1 nonplanar diagram includes an ultraviolet divergence,
we have to renormalize it by adding a contribution
from a diagram involving
the one-loop counterterm for the three-point function
[ diagram (o) in Fig.\ \ref{fig:two-point} ].
%(See Fig.\ \ref{nonplanar1}.)
After this procedure, we can study the behavior 
at infinite external momentum.
We adopt dimensional regularization and
take the space-time dimensionality to be $d=6-\varepsilon$.
Because the coupling constant $g$ has dimensions $(\rm{mass})^{(6-d)/2}$, 
we set $g=\mu^{\frac{\varepsilon}{2}} g_r$, where
$\mu$ is the renormalization point, and $g_r$ is a
dimensionless coupling constant.
The ultraviolet divergence appears as a $1/\varepsilon$ pole
in the $d \rightarrow 6$ limit.

\begin{figure}[htb]
\begin{center}
\includegraphics[width = 10cm,clip]{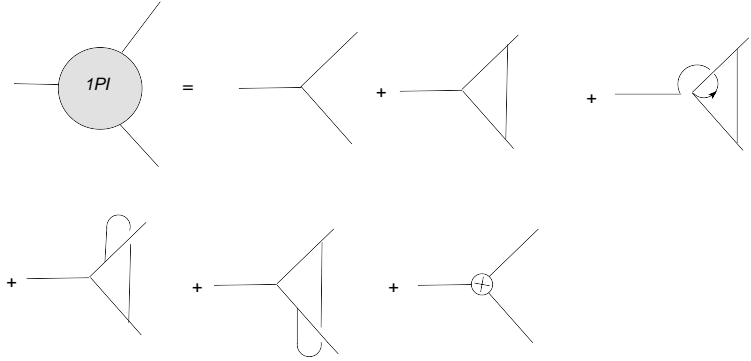}
\end{center}
\caption{The list of diagrams needed to calculate the three-point function
up to the one-loop level.}
\label{thp}
\end{figure}

%\begin{wrapfigure}{l}{6.6cm}
%\includegraphics[width = 3cm,clip]{vertex.eps}
%\caption{The planar one-loop diagram for calculating $\Gamma(p, q)$.}
%\label{vertex}
%\end{wrapfigure}

\begin{figure}[htb]
\begin{flushleft}
\begin{center}
\includegraphics[width = 3cm,clip]{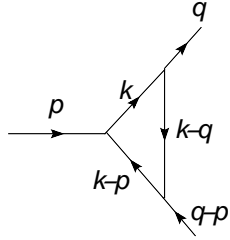}
\end{center}
\end{flushleft}
\caption{The planar one-loop diagram for calculating $\Gamma(p, q)$.}
\label{vertex}
\end{figure}

The one-loop counterterm for the three-point function
can be determined in such a way that the three-point function
becomes finite at the one-loop level.
The relevant diagrams are listed in Fig.\ \ref{thp}.
Because the nonplanar diagrams are finite due to the insertion
of the momentum dependent phase factor,
we only need to calculate the planar diagram depicted in
Fig.\ \ref{vertex}, which can be evaluated as
%Assigning momentum variables as in Fig.\ \ref{vertex}, we calculate
%the one-loop diagram as
\beqa
 \Gamma(p, q) & =&  g^3 \int \frac{d^{d}k}{(2\pi)^d}
\frac{1}{k^2 + m^2}\, \frac{1}{(k-q)^2 + m^2}\,
\frac{1}{(k-p)^2 + m^2} \nonumber  \\
 &=& \frac{g^3}{(4\pi)^\frac{d}{2}} 
\int_{0}^{1}d \alpha\, \alpha^{\frac{d}{2}-3} (1- \alpha )^{\frac{d}{2}-2}
\int_{0}^{1}d \beta \int_{0}^{\infty}dt\, 
\frac{1}{t^{\frac{d}{2}-2}}\exp( -t \Delta )  \ ,
\label{Gamma-1-loop}
\eeqa
where $\Delta$ is defined by
\beq
\Delta  =  \frac{m^2}{\alpha(1-\alpha )} + q^2 - 2\beta\, p \cdot q
+ \frac{\beta}{\alpha} \Bigl( 1-(1-\alpha) \beta \Bigr ) \, p^2  \ .
\eeq
The divergent part can be extracted as
\beq
 \Gamma(p, q)
= \frac{g_r^3}{(4\pi)^3}\,\frac{1}{\varepsilon} 
+ \mbox{O}(\varepsilon ^0) \ ,
\eeq
%Taking account of the same contribution coming from an analogous diagram, 
from which we determine the one-loop counter-term as
\beq
 \delta_g  = -\frac{g_r^3}{(4\pi)^3} \, 
\frac{ \mu^{\frac{\varepsilon}{2}}}{\varepsilon} \ .
\label{delta-g}
\eeq

\begin{figure}[htb]
\begin{center}
\includegraphics[width = 5cm,clip]{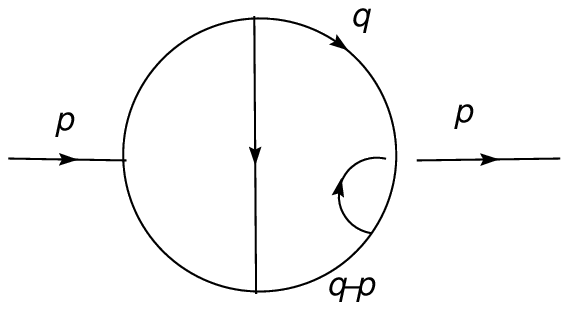}
\end{center}
\caption{The type 1 nonplanar diagram for calculating 
$\Pi_{\rm NP1}( p^2)$.}
\label{type1}
\end{figure}

Using Eq.\ (\ref{Gamma-1-loop}),
we evaluate the type 1 nonplanar diagram 
in Fig.\ \ref{type1} as
\beqa
 \Pi_{{\rm NP}1}( p^2) 
&= & g \int \frac{d^{d}q}{(2\pi)^d}\, \Gamma(p, q)\, \frac{1}{q^2 + m^2}\,
\frac{1}{(q-p)^2 + m^2}\, e^{i q \cdot \theta p}
\nonumber  \\
& = & \frac{g^4}{(4\pi)^d} \biggl[ \frac{4}{(\theta p)^2} \biggr ]^{d-5} 
\int_{0}^{1}d \alpha\, \alpha^{\frac{d}{2}-3} 
(1-\alpha)^{\frac{d}{2}-2} \int_{0}^{1}d \beta \int_{0}^{1}
d \zeta\, \zeta^{2-\frac{d}{2}} (1- \zeta)  \nonumber \\
&~ &  \cdot \int_{0}^{1} d \eta
\int_{0}^{\infty} dt\, t^{d-6} 
\exp \Bigl ( -t- \frac{\tilde{\Delta }}{4t}\, p^2 ( \theta p)^2 \Bigr ) \ ,
\label{PiNP1}
\eeqa
where $(\theta p)^2 \equiv
-p^{\mu }\theta ^{\mu \nu }\theta ^{\nu \rho }p^\rho \geq 0$,
since $\theta ^{\mu \nu }$ is an antisymmetric tensor.
The quantity $\tilde{\Delta}$ is defined as
\beq
\tilde{\Delta } = 
\frac{m^2}{p^2} 
\biggl[ \frac{\zeta}{\alpha ( 1-\alpha) } + 1 - \zeta  \biggr ] + \,
\frac{1}{\alpha} 
\biggl [\zeta\, \beta (1-\beta) + 
\alpha (1-\zeta) \Bigl( \zeta(\beta - \eta)^2
+ \eta (1- \eta) \Bigr) \biggr] \ .
\eeq
Extracting the divergent part of (\ref{PiNP1}) and 
taking the $d \rightarrow 6$ limit for the finite part,
we obtain
%%\beqa
%%\Pi_{{\rm NP}1}( p^2) 
%%& = & \frac{g_r^4}{(4\pi)^6}\,\frac{4}{(\theta p)^2}\,
%%\frac{1}{\varepsilon}  \mathcal{A }\Bigl( \frac{m^2}{p^2};\, 
%%p^2 (\theta p)^2 \Bigr )
%%\nonumber \\
%%&~& + \frac{g_r^4}{(4\pi)^6}\,
%%\frac{4}{(\theta p)^2}\biggl[ \Bigl ( \ln\,( \pi \mu^2 (\theta p)^2) + 1 
%%\Bigr ) \mathcal{A}\Bigl( \frac{m^2}{p^2};\, p^2 (\theta p)^2 \Bigr )
%%\nonumber \\
%%& ~& - \mathcal{B} \Bigl( \frac{m^2}{p^2};\, p^2 ( \theta p)^2 \Bigr )
%%-\mathcal{C} \Bigl( \frac{m^2}{p^2};\, p^2( \theta p)^2 \Bigr ) \biggr ]
%%\nonumber \\
%%& ~& + \frac{g_r^4}{(4\pi)^6}\, p^2  \mathcal{D} \Bigl( \frac{m^2}{p^2};\, 
%%p^2 ( \theta p)^2 \Bigr ) \ , 
%%\label{PI-NP-1}
%%\eeqa
%%where we have introduced
\beq
\Pi_{{\rm NP}1}( p^2) 
= \frac{g_r^4}{(4\pi)^6}
\left[
%\frac{4}{(\theta p)^2}\,
%\frac{\mathcal{A }}{\varepsilon}\,  
% \frac{g_r^4}{(4\pi)^6}\,
%+
\frac{4}{(\theta p)^2}\biggl\{ 
\frac{\mathcal{A }}{\varepsilon} +
\Bigl ( \ln\,( \pi \mu^2 (\theta p)^2) + 1 
\Bigr ) \,\mathcal{A}
%\nonumber \\
- \,\mathcal{B}
-\,\mathcal{C} \biggr\}
%\Bigl\}
%\nonumber \\
+ 
%\frac{g_r^4}{(4\pi)^6}
p^2 \mathcal{D} \right] \ ,
\label{PI-NP-1}
\eeq
where we have introduced
%the $\mathcal{A,B,C,D}$ that are functions of 
%$\frac{m^2}{p^2};p^2(\theta p^2)$ as follows. 
%
%% \begin{align}
%% &\mathcal{A} \Bigl( \frac{m^2}{p^2};\, p^2 ( \theta p)^2 \Bigr ) = \int_{0}^{1} d \eta
%% \int_{0}^{\infty} dt\, \exp \biggl[ -t - \frac{\frac{m^2}{p^2} + 
%% \eta (1-\eta)}{4t}\, p^2 ( \theta p)^2 \biggr ] \ ,
%% \label{ANC}  \\
%% &\mathcal{B} \Bigl( \frac{m^2}{p^2};\, p^2 ( \theta p)^2 \Bigr ) = \int_{0}^{1} d \eta
%% \int_{0}^{\infty} dt\,(\ln t) \exp \biggl[ -t - \frac{\frac{m^2}{p^2} + \eta (1-\eta)}{4t}\, 
%% p^2 ( \theta p)^2  \biggr ] \, \label{BNC}  \\
%% &\mathcal{C} \Bigl( \frac{m^2}{p^2};\, p^2 ( \theta p)^2 \Bigr ) = \int_{0}^{1} d \alpha 
%% \int_{0}^{1} d \beta \int_{0}^{1} d \zeta \int_{0}^{1}d \eta  \int_{0}^{\infty} dt\,(1-\alpha)
%% \exp \biggl[ -t - \frac{\tilde{\Delta }}{4t}\, p^2 ( \theta p)^2 \biggr ] \ ,
%% \label{CNC}  \\
%% &\mathcal{D} \Bigl( \frac{m^2}{p^2};\, p^2 ( \theta p)^2 \Bigr ) = \int_{0}^{1} d \alpha \int_{0}^{1} d \beta
%% \int_{0}^{1} d \zeta \int_{0}^{1}d \eta \int_{0}^{\infty} dt\, 
%% \frac{F \ln \zeta}{\alpha t}
%% \exp \biggl[ -t - \frac{\tilde{\Delta }}{4t}\, p^2 ( \theta p)^2 \biggr ] \ .
%% \label{DNC} 
%% \end{align}
\beqa
\mathcal{A} &=& \int_{0}^{1} d \eta
\int_{0}^{\infty} dt\, \exp \biggl[ -t - \frac{\frac{m^2}{p^2} + 
\eta (1-\eta)}{4t}\, p^2 ( \theta p)^2 \biggr ] \ ,
\label{ANC}  \\
\mathcal{B} &=& \int_{0}^{1} d \eta
\int_{0}^{\infty} dt\,(\ln t) \exp 
\biggl[ -t - \frac{\frac{m^2}{p^2} + \eta (1-\eta)}{4t}\, 
p^2 ( \theta p)^2  \biggr ], \, \label{BNC}  \\
\mathcal{C} &=& \int_{0}^{1} d \alpha 
\int_{0}^{1} d \beta \int_{0}^{1} d \zeta 
\int_{0}^{1}d \eta  \int_{0}^{\infty} dt\,(1-\alpha)
\exp \biggl[ -t - \frac{\tilde{\Delta }}{4t}\, p^2 ( \theta p)^2 \biggr ] \ ,
\label{CNC}  \\
\mathcal{D} &=& \int_{0}^{1} d \alpha \int_{0}^{1} d \beta
\int_{0}^{1} d \zeta \int_{0}^{1}d \eta \int_{0}^{\infty} dt\, 
\frac{F \ln \zeta}{\alpha t}
\exp \biggl[ -t - \frac{\tilde{\Delta }}{4t}\, p^2 ( \theta p)^2 \biggr ] \ ,
\label{DNC} 
\eeqa
which are functions of $\frac{m^2}{p^2}$ and $p^2(\theta p)^2$.
The coefficient $F$ in Eq.\ (\ref{DNC})
is defined by
\beq
 F = \frac{m^2}{p^2}\Bigl(1-\alpha(1-\alpha) \Bigr)
+ (1-\alpha) \biggl[ \beta(1-\beta) + \alpha
\Bigl( (1-2\zeta)(\beta - \eta)^2 - \eta(1-\eta) \Bigr ) \biggr] \ . 
\label{FNC}
\eeq

We now demonstrate that
the functions $\mathcal{A}$, $\mathcal{B}$, 
$\mathcal{C}$ and $\mathcal{D}$ vanish in the $p^2 \to \infty$ limit. 
For this purpose,
we first confirm the convergence of all the integrals.
Then it will suffice to show that the integrands vanish
in the $p^2 \to \infty$ limit.
%% When the fact \cite{rf:MRS} was confirmed that 
%% the maximally non-commutative theory 
%% which is obtained in the limit $\theta \to \infty$
%% is given by the planar diagrams only, such procedure was taken.
The convergence of the integrals 
in $\mathcal{A}, \mathcal{B}$ and $\mathcal{C}$ is
evident. As for the function $\mathcal{D}$,
the $\alpha$- and $t$-integrals would have
singularities at the lower ends of the integration domain
if $\theta_{\mu\nu}$ were zero, but they are regularized 
by the term proportional to $1/(\alpha t)$ in the exponent,
which appears for nonzero $\theta_{\mu\nu}$.
Actually, we can put an upper bound on 
the absolute values of these functions
by integrating elementary functions 
that are larger than the corresponding integrands. 
In this way, we obtain upper bounds on
$|\mathcal{A}|$, $|\mathcal{B}|$ and $|\mathcal{C}|$ as
\beq
 |\mathcal{A} | < 1 \ , 
\quad |\mathcal{B}|<2 \ , 
\quad |\mathcal{C}| < 1 \ ,  \label{ABCbound} 
\eeq
by omitting the term proportional to $p^2 ( \theta p)^2$ in the exponent.
Obtaining an upper bound on $|\mathcal{D}|$ is more involved,
due to the ``noncommutative'' regularization 
of the singularities mentioned above,
%due to the cancellation of the sin
but the calculation in Appendix \ref{sec:derive-ub} yields
\beq
|\mathcal{D}| < 
\frac{32}{p^2 ( \theta p)^2}\Bigl( 1 + \ln\frac{p^2}{m^2} \Bigr)
\Bigr( 4 + \ln\frac{p^2}{m^2} + \ln\frac{ p^2 ( \theta p)^2}{4} \Bigr) \ ,
\label{upper-bound}
\eeq
where we have assumed $m^2/p^2 \ll 1$, since we are ultimately interested
in the $p^2 \rightarrow \infty$ limit.
%(The upper bounds (\ref{ABCbound}) can be obtained by
%omitting the term proportional to $p^2 ( \theta p)^2$ in the exponent.
%See Appendix \ref{sec:derive-ub} for the derivation of
%eq.\ (\ref{upper-bound}).)
This confirms the convergence.
Because the integrands of the functions 
$\mathcal{A}$, $\mathcal{B}$, 
$\mathcal{C}$ and $\mathcal{D}$ 
decrease exponentially at large $p^2$,
we conclude that all the functions vanish 
in the $p^2 \to \infty$ limit.

\begin{figure}[htb]
%\begin{flushleft}
\begin{center}
\includegraphics[width = 4cm,clip]{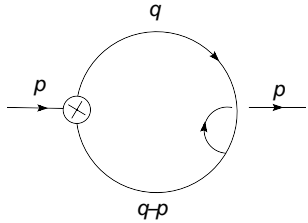}
%\end{flushleft}
\end{center}
\caption{The nonplanar one-loop
diagram for calculating $\Pi_{{\rm NP1,ct}}(p^2)$.
%including the counter-term for the three-point vertex.
}
\label{type1c}
\end{figure}

Next we evaluate the diagram in Fig.\ \ref{type1c} 
involving the counterterm (\ref{delta-g}) as
%, which involves the counter-term (\ref{delta-g}), as
%% \beqa
%% &~& \Pi_{{\rm NP}1, {\rm ~ct}}( p^2)  \nonumber \\
%% & = & \delta_g\, g\, 
%% \int \frac{d^{d}k}{(2\pi)^d} \frac{1}{k^2 + m^2}\,
%% \frac{1}{(k-p)^2 + m^2} 
%% e^{i k\cdot\, \theta p}
%% \nonumber \\
%% & = &  \delta_g\, \frac{g}{(4\pi)^{\frac{d}{2}}}
%% \biggl [ \frac{4}{( \theta p)^2} \biggr ] ^{\frac{d}{2}-2}
%% \int_{0}^{1} d \alpha\, \int_{0}^{\infty}dt \,
%% t^{\frac{d}{2}-3}\, \exp \biggl[ -t - \frac{\frac{m^2}{p^2} 
%% + \alpha (1-\alpha)}{4t}\, p^2 ( \theta p)^2 \biggr ]
%% \nonumber  \\
%% & = & -\frac{4 g_r^4}{(4\pi)^6}\, \frac{1}{( \theta p)^2}\,
%% \frac{1}{\varepsilon}
%% \mathcal{A} \Bigl( \frac{m^2}{p^2};\, p^2 ( \theta p)^2 \Bigr )
%% \nonumber  \\
%% &~&  
%% + \frac{4 g_r^4}{(4\pi)^6}\, \frac{1}{( \theta p)^2}
%% \biggl[-\frac{1}{2}\ln\,( \pi \mu^2 ( \theta p)^2)\,
%% \mathcal{A} \Bigl( \frac{m^2}{p^2};\, p^2 ( \theta p)^2 \Bigr ) 
%% + \,\frac{1}{2}
%% \mathcal{B} \Bigl( \frac{m^2}{p^2};\, p^2 ( \theta p)^2 \Bigr ) \biggr] \ .
%% \label{NP-ct}
%% \eeqa
\beqa
\Pi_{{\rm NP1,ct}}(p^2) 
& = & \delta_g\, g\, 
\int \frac{d^{d}k}{(2\pi)^d} \frac{1}{k^2 + m^2}\,
\frac{1}{(k-p)^2 + m^2}  \,  e^{i k\cdot\, \theta p}
\nonumber \\
& = &  \delta_g\, \frac{g}{(4\pi)^{\frac{d}{2}}}
\biggl [ \frac{4}{( \theta p)^2} \biggr ] ^{\frac{d}{2}-2}
\int_{0}^{1} d \alpha\, \int_{0}^{\infty}dt \,
t^{\frac{d}{2}-3}\, \exp \biggl[ -t - \frac{\frac{m^2}{p^2} 
+ \alpha (1-\alpha)}{4t}\, p^2 ( \theta p)^2 \biggr ]
\nonumber  \\
& = & \frac{4 g_r^4}{(4\pi)^6}\, \frac{1}{( \theta p)^2}\,
\left[ - \frac{\mathcal{A}}{\varepsilon}
+ \biggl\{ -\frac{1}{2}\ln\,( \pi \mu^2 ( \theta p)^2)\,
 \mathcal{A} + \,\frac{1}{2} \, \mathcal{B} \biggr\} \right]
\ .
\label{NP-ct}
\eeqa
The first term cancels the $1/\varepsilon$
pole in (\ref{PI-NP-1}),
and we have taken the $d \rightarrow 6$ limit
for the remaining terms.
%% However there is a singular factor $1/(\theta p)^2$ 
%% in the commutative limit $\theta \to 0$ 
%% as in the previous diagram. 
%% Therefore the first term can be identified
%% as the double pole and the other terms
%% as the simple poles in the commutative case.

Adding (\ref{PI-NP-1}) and (\ref{NP-ct}),
we obtain the finite result 
%% \beqa
%% \Pi_{{\rm NP1r}}(p^2) &=&
%% \frac{g_r^4}{(4\pi)^6}\,
%% \frac{2}{( \theta p)^2}\biggl[ \Bigl ( \ln\,( \pi \mu ^2 ( \theta p)^2)
%% + 2 \Bigr ) A \Bigl( \frac{m^2}{p^2};\, p^2 ( \theta p)^2 \Bigr )\,-
%% B \Bigl( \frac{m^2}{p^2};\,p^2 ( \theta p)^2 \Bigr )
%% \nonumber \\
%% &~& -2C \Bigl( \frac{m^2}{p^2};\, p^2 ( \theta p)^2 \Bigr ) \biggr ]\,  
%% + \frac{g_r^4}{(4\pi)^6}\, p^2
%% D \Bigl( \frac{m^2}{p^2};\, p^2 ( \theta p)^2 \Bigr ) \ .
%% \label{Pi-ren}
%% \eeqa
\beq
\Pi_{{\rm NP1r}}(p^2) =
\frac{g_r^4}{(4\pi)^6}
\left[
\frac{2}{( \theta p)^2}\biggl\{ \Bigl ( \ln\,( \pi \mu ^2 ( \theta p)^2)
 + 2 \Bigr ) \mathcal{A}\,-\mathcal{B} -2\,\mathcal{C} \biggr \} 
+ 
% \frac{g_r^4}{(4\pi)^6}\, 
p^2\mathcal{D} \right]  \ .
\label{Pi-ren}
\eeq
Because the functions $\mathcal{A}$, $\mathcal{B}$, 
$\mathcal{C}$ and $\mathcal{D}$ 
vanish in the $p^2 \to \infty$ limit, 
as shown above, the 
type 1 nonplanar diagram, 
% $\Pi_{{\rm NP1r}}(p^2)$ 
after renormalizing the divergence from the planar
subdiagram, vanishes in the same limit.
In fact, using (\ref{ABCbound}) and (\ref{upper-bound}),
%in eq.\ (\ref{Pi-ren}), 
we obtain an upper bound on $|\Pi_{{\rm NP1r}}(p^2)|$ as
\beqa
\left| \Pi_{{\rm NP1r}}(p^2) \right| 
&<& \frac{8 g_r^4}{(4\pi)^6}\frac{1}{( \theta p)^2} \nonumber \\
&~& \cdot \biggl[ \frac{3}{2} 
+ \frac{1}{4}\ln ( \pi \mu ^2 ( \theta p)^2 ) 
+ 4\Bigl( 1 + \ln\frac{p^2}{m^2} \Bigr)
\Bigr( 4 + \ln\frac{p^2}{m^2} + 
\ln\frac{p^2 ( \theta p)^2}{4} \Bigr) \biggr] \ ,
\nonumber
\eeqa
where the right-hand side does vanish in the 
$p^2 \rightarrow \infty$ limit,
thus confirming the above conclusion more explicitly.
%, although the l.h.s.\ actually vanishes much faster (exponentially).
%
%% Thus we conclude
%% that the nonplanar diagram which includes the $\log$ divergence 
%% vanishes in the $p^2 \to \infty$ limit 
%% after we renormalized the divergences.

\section{Type 2 diagram}
\label{section:type2}

\begin{figure}[htb]
%\begin{flushleft}
\begin{center}
\includegraphics[width = 3cm,clip]{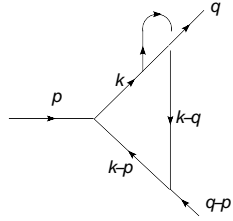}
%\end{flushleft}
\end{center}
\caption{The nonplanar one-loop diagram
for calculating $\Gamma_{\rm NP}(p,q)$.
%vertex correction at the one-loop level. 
}
\label{nonvertex}
\end{figure}

In this section, we consider the type 2 nonplanar diagram,
in which internal lines cross, and 
study its behavior in the $p^2 \to \infty$ limit.
Let us first 
evaluate the one-loop subdiagram in Fig.\ \ref{nonvertex}. We have
\beqa
\Gamma_{\rm NP}(p,q) 
&=& 
g^3 \int \frac{d^{d}k}{(2\pi)^d}\frac{1}{k^2 + m^2}\, 
\frac{1}{(k-q)^2 + m^2}\,
\frac{1}{(k-p)^2 + m^2} \, e^{i k \cdot \theta q} 
\nonumber \\
&=& \frac{g^3}{(4\pi)^\frac{d}{2}} 
\int_{0}^{1}d \alpha\, \alpha^{\frac{d}{2}-3} (1- \alpha )^{\frac{d}{2}-2}
\int_{0}^{1}d \beta \int_{0}^{\infty}dt\, \frac{1}{t^{\frac{d}{2}-2}}
\nonumber \\
&~& \cdot \exp\biggl[ -t\Delta - \frac{\alpha (1-\alpha )}{4t}\,
( \theta q)^2 + i\,\beta\, (1-\alpha )\,
p\theta \cdot q \biggr]  
\label{Gamma-NPbefore}
\\
&=& \frac{2 g_{r}^3}{(4\pi )^3}\int_{0}^{1}d\alpha \,  (1-\alpha )
\int_{0}^{1}d\beta\, e^{i\beta (1-\alpha ) p\theta \cdot q} 
K_0 \left(\sqrt{\alpha (1-\alpha ) \Delta ( \theta q)}\right)  \ 
\label{Gamma-NP}
\eeqa
where $K_0(z)$ is the modified Bessel function, and 
$\Delta$ in (\ref{Gamma-NPbefore}) is defined by
\beq
\Delta  =   \frac{m^2}{\alpha(1-\alpha )}
+ q^2 - 2\beta\, p \cdot q + 
\frac{\beta}{\alpha} \Bigl( 1-(1-\alpha) \beta \Bigr ) p^2 \ .
\eeq
In proceeding from (\ref{Gamma-NPbefore}) to (\ref{Gamma-NP}),
we have taken the $d \to 6$ limit.
This diagram is finite, as the logarithmic ultraviolet divergence, 
which would arise in the commutative case,
is regularized by the non-commutative phase factor.
We can understand this fact by considering the asymptotic behavior 
of (\ref{Gamma-NP})
\beq
\Gamma_{\rm NP}(p,q) 
\simeq  - \frac{g_{r}^3}{2(4\pi )^3} \ln p^2 ( \theta q)^2 
%\quad \quad (p^2 \neq 0,\, (\theta q)^2 \approx 0 ) 
\label{Gam-NP-theta0}
\eeq
for $(\theta q)^2 \approx 0 $ at nonzero $p^2$.
This logarithmic behavior reflects the
ultraviolet divergence in the commutative case.
This can be regarded as a result of the UV/IR mixing.

%% Next we calculate the two-loop nonplanar diagram,
%% which includes the vertex (\ref{Gamma-NP}).
%% Though it is our purpose that 
%% we observe how this diagram behaves in the limit $p^2 \to \infty$, 
%% we are also interested in
%% what occurs on the infrared singularity (\ref{Gam-NP-theta0})
%% in the lower limit of the integral over the internal momentum $q^2$.
%% Regarding this interest, we will have found that 
%% this singularity vanishes after integrating over $q^2$.

In what follows,
we assume for simplicity
that all the eigenvalues of the symmetric matrix 
$( \theta ^2 )^{\mu \rho} = \theta ^{\mu \nu }\theta ^{\nu \rho }$ 
are equal, and denote it as $-\theta ^2\,\,( < 0 ) $. 
The general case is considered later. 
%We will see that the behavior in the $p^2 \to \infty$ limit 
%is qualitatively the same as far as 
%all the eigenvalues are nonzero.
In fact, when some of the eigenvalues are zero,
% on the other hand,
there are certain differences in the 
behavior at large $p^2$, but
our final conclusion concerning 
the $p^2 \rightarrow \infty$ limit is the same.

%% For example, there are only two non zero eigenvalues in the case 
%% that there is only non-commutativity of 
%% 1,2-direction in d-dimentional space. 
%% And there are four non zero eigenvalues in the case that
%% there are non-commutativities of 1,2,3,4-direction. 

\begin{figure}[htb]
%\begin{flushleft}
\begin{center}
\includegraphics[width = 5cm,clip]{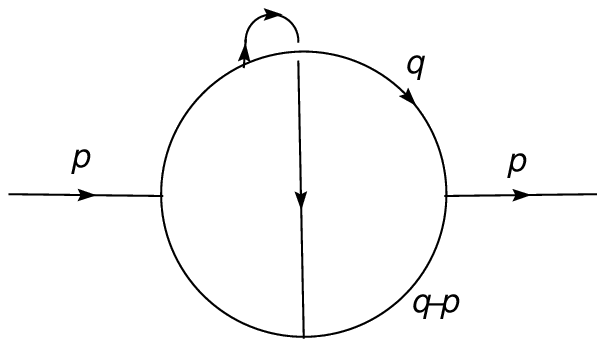}
%\end{flushleft}
\end{center}
\caption{The type 2 nonplanar diagram for 
calculating $\Pi_{\rm NP2}( p^2)$.
}
\label{type2}
\end{figure}

Using (\ref{Gamma-NPbefore}), we evaluate the diagram 
in Fig.\ \ref{type2} as
\beqa
&~& \Pi_{\rm NP2}( p^2)   \nonumber   \\  
& = & g \int \frac{d^{d}q}{(2\pi)^d}\, 
\Gamma_{\rm NP}(p, q) \frac{1}{q^2 + m^2}\,
\frac{1}{(q-p)^2 + m^2}
\nonumber  \\
& = & \frac{g^4}{(4\pi)^d}(p^2)^{d-5}
\int_{0}^{1}d\alpha\, \alpha ^{\frac{d}{2}-3}(1-\alpha )^{\frac{d}{2}-2}
\int_{0}^{1}d\beta \int_{0}^{1}d\zeta\, 
\zeta ^2(1-\zeta )\int_{0}^{1}d\eta
\nonumber \\
&~& \cdot \int_{0}^{\infty }dt\, t^4
\frac{1}{\Bigl[t^2 \zeta + \frac{\alpha (1-\alpha )}{4}\,
\theta ^2 (p^2)^2 \Bigr]^{\frac{d}{2}}}
\exp \biggl[ -t\tilde\Delta - t 
\frac{\tilde\Delta_{\rm NP}}{t^2 \zeta + 
\frac{\alpha (1-\alpha )}{4}\,\theta ^2 (p^2)^2}  \biggr] \ ,
\label{Pi-NP2}
\eeqa
where we have defined

\beqa
\tilde{\Delta } & \equiv  & 
\frac{m^2}{p^2} 
\biggl[ \frac{\zeta}{\alpha ( 1-\alpha) } + 1 - \zeta  \biggr ] + 
\frac{\zeta \beta (1-\beta)}{\alpha} 
+ (1-\zeta) 
\Bigl( \zeta(\beta - \eta)^2
+ \eta (1- \eta) \Bigr)  \ , \mbox{~~~~}
\label{tildeDelta}
\\
\tilde{\Delta }_{\rm NP} & \equiv & 
\frac{ 1-\alpha }{4} \Bigl[ \zeta \beta ^2(1-\alpha ) 
+ \alpha \bigr(\zeta \beta + 
(1-\zeta )\eta \bigl)^2 \Bigr]\theta ^2 (p^2)^2 \ .
\label{tildeDeltaNP}
\eeqa
%First let us confirm the convergence of each integral.
The $\zeta$- and $t$-integrals would have
singularities at the lower ends of the integration domain
if $\theta^2$ were zero, but they are regularized 
by the term proportional to $\theta ^2$ in the denominator.
This appearance of the $\theta ^2$ term is 
peculiar to nonplanar diagrams 
in which the internal lines cross \cite{rf:MRS}.
By omitting the $t^2 \zeta$ term in the denominator
and the terms other than those proportional to $m^2$
in the exponent,
we obtain an upper bound on $|\Pi_{\rm NP2}( p^2)|$ as
\footnote{
Because the upper bound (\ref{upper-bound2})
is independent of $p^2$,
we find that $|\Pi_{\rm NP2}( p^2)|$ is finite
even in the $p^2 \rightarrow 0$ limit.
This is in contrast to the type 1 diagram (after renormalization), 
which actually diverges
in the $p^2 \rightarrow 0$ limit, due to the UV/IR mixing.}
\beq
|\Pi_{\rm NP \, 2}( p^2)| 
< \frac{g^4}{(4\pi)^d}\frac{4^{\frac{d}{2}}}{(\theta ^2)^{\frac{d}{2}}
(m^2)^5} \ ,
\label{upper-bound2}
\eeq
which confirms the convergence of the multiple integral 
in (\ref{Pi-NP2}) in the $d \to 6$ limit.
Given this, the fact that 
the integrand in Eq.\ (\ref{Pi-NP2})
decreases as $1/(p^2)^5$ at large $p^2$
implies that the type 2
nonplanar diagram vanishes in the $p^2 \to \infty $ limit.
%on the grounds of the convergence demonstrated above.
As we have done in the case of type 1 diagrams, we can actually 
put a more stringent upper bound on $|\Pi_{\rm NP2}( p^2)|$, indeed, one
which vanishes in the $p^2 \to \infty $ limit.
This confirms our assertion more explicitly.
(See Appendix \ref{sec:derive-ub2} for the details.)

Let us comment on the case in which the number of 
non-commutative directions
is less than the space-time dimensionality, $d$.
In this case,
the upper bound on $|\Pi_{\rm NP2}( p^2)|$ can be evaluated as
\beqa
| \Pi_{\rm NP2}( p^2) | & < & 
\frac{g^4}{(4\pi)^d}(p^2)^{d-5}
\int_{0}^{1}d\alpha\, \alpha ^{\frac{d}{2}-3}(1-\alpha )^{\frac{d}{2}-2}
\int_{0}^{1}d\beta \int_{0}^{1}d\zeta\, \zeta ^2(1-\zeta )
\int_{0}^{1}d\eta 
\nonumber \\
&~& \cdot \int_{0}^{\infty }dt\, t^4
\prod_{j = 1}^{d}\frac{1}{\Bigl[t^2 \zeta + 
\frac{\alpha (1-\alpha )}{4}\,\theta _{j}^2 (p^2)^2 \Bigr]^{\frac{1}{2}}}
\exp \Bigl[ -t\tilde\Delta \Bigr] \ ,
\label{upper-bound3}
\eeqa
where $(-\theta _{j}^2)$ is the $j$-th eigenvalue 
of $(\theta ^2)^{\mu \nu }$, 
and $\tilde\Delta$ is the same as that defined in (\ref{tildeDelta}).
If non-commutativity is introduced in only $k$ directions
$(2\leq k\leq 6)$,
we obtain an upper bound from (\ref{upper-bound3}),
generalizing (\ref{upper-bound2}), as
\begin{equation}
| \Pi_{\rm NP2}( p^2) | 
< \frac{g^4}{(4\pi)^d}\frac{a_{k}\cdot  4^{\frac{k}{2}}}
{\prod_{j = 1}^{k} (\theta _{j} ^2)^{\frac{1}{2}}(m^2)^{k-1}} \ , 
\end{equation} 
where $a_{k}$ is a $k$-dependent constant
that is irrelevant to the issue with which we are presently concerned.
In the $p^2 \to \infty $ limit,
the integrand on the right-hand side of (\ref{upper-bound3})
decreases as $(1/p^2)^{k-1}$.
Thus we conclude that the type 2 nonplanar diagram vanishes 
in the $p^2 \to \infty$ limit for general $\theta_{\mu\nu}$.
%% the more the number of non-commutative
%% directions is,
%% the faster the type 2 nonplanar diagram vanishes 
%% in the $p^2 \to \infty$ limit.

\section{Summary and discussion}
\label{section:summary}

In this paper we have studied
the vanishing of nonplanar diagrams in the $p^2 \to \infty $ limit
in 6d non-commutative $\phi ^3$ theory at the two-loop level.
We have investigated two types of nonplanar diagrams separately.
In the type 1 nonplanar diagram,
%in which the external line and the internal line cross,
we have confirmed the planer dominance after
renormalizing the ultraviolet divergence
coming from the planar subdiagram.
In the type 2 nonplanar diagram
the planer dominance holds despite the fact that
the non-commutative phase factor does not depend on the external
momentum. 

Based on the behavior observed for these two types of diagrams,
we can argue that the other types of nonplanar diagrams 
in Fig.\ \ref{fig:two-point} also vanish in the $p^2 \to \infty $ limit.
The situations for the diagrams (k) and (l) are analogous 
to those for the type 1 and type 2 diagrams, respectively.
The diagram (k) has a planar subdiagram, which causes a UV
divergence. This divergence can be cancelled by adding the contribution
from the diagram (q), and the resulting finite quantity should vanish
due to the crossing of an external line and an internal line.
The diagram (l) is finite by itself, and it should vanish in a 
manner similar to the type 2 diagram due to the crossing of
internal lines.
The diagrams (g),(h) and (i) have more crossings of momentum lines
than the type 1 diagram. Therefore for those diagrams, there are extra non-commutative
phase factors, which make the diagrams finite by themselves. 
The vanishing of these diagrams then follows as in the case of the type 1 diagram.
%% depends only on internal momenta.
%% This leads to the qualitatively different behaviors.
%, in which the internal lines cross,
%% vanishes as $(1/p^2)^{k-1}$ in the $p \to \infty $ limit,
%% where $k$ represents the number of non-commutative directions.
An analogous argument applies to the diagram (m), which has more crossings
of momentum lines than the diagram (k).
%, which make it finite by itself and the vanishing
%occurs as in the diagram (k).
%
Although we have studied a particular model for concreteness, 
we believe that the same conclusion holds for a more general class of models.

We should mention that the renormalization procedure 
\cite{Chepelev:1999tt}
in non-commutative scalar field theories
encounters an obstacle due to severe infrared divergence at higher loops
\cite{rf:MRS}.
This problem can be overcome by resumming a class of diagrams
with infrared divergence in $\phi^4$ theory \cite{rf:MRS}.
Indeed, Monte Carlo simulations
% of the corresponding lattice theory
%non-commutative $\phi^4$ theory
%on the lattice 
show that one can obtain a sensible 
continuum limit \cite{Bietenholz:2004xs}, which suggests
the appearance of a dynamical infrared cutoff due to 
{\em nonperturbative} effects.
Introducing an infrared cutoff with such a dynamical origin
in perturbation theory,
we believe that the result obtained here up to two-loop order
can be generalized to all orders. 
%
%We expect that the statement we have confirmed should hold 
%beyond perturbation theory.

%% Looking at the results, 
%% the effect of $p \to \infty $ is almost the same 
%% as one of $\theta \to \infty $.
%% We can understand the meaning of this phenomenon as follows. 
%% There are some places where the $\theta $ and $p$
%% appear in the two point functions 
%% which we have calculated in momentum space.
%% First there is necesarily the overall factor $p^2$, 
%% because the diagrams which presented 
%% the parturbative effect have mass dimension 2.  
%% Secondly in the exponent the $\theta $ and $p$ 
%% appear in the combination $ \theta ^2(p^2)^2$ that has no dimension. 
%% Thirdly the terms of infrared singularity appear 
%% in the $1/\theta ^2(p^2)^2$, or $ \ln\theta ^2(p^2)^2 $.
%% We therefore can see that $\theta $ and $p$ appear 
%% side by side except the overall factor $p^2$. 

\section*{Acknowledgements}
We would like to thank S.~Iso and H.~Kawai for fruitful discussions.
The work of J.~N.\ 
is supported in part by a Grant-in-Aid for 
Scientific Research (No.\ 14740163)
from the Ministry of Education, Culture, Sports, Science 
and Technology of Japan. 

\bigskip

\appendix

%%%%%%%%%%%%%% added by F.S. 2003.3.3 (beginning) %%%%%%%%%%%%%%%%%%%%%%
\section{Derivation of the Upper Bound (\ref{upper-bound})}
\label{sec:derive-ub}
\setcounter{equation}{0}
\renewcommand{\theequation}{A$\cdot$\arabic{equation}}
In this appendix we derive the upper bound (\ref{upper-bound}) under the condition $\frac{m^2}{p^2}\ll 1$.
Let us consider 
$\tilde{\Delta }$ and $F$ in Eq.\ 
(\ref{DNC}). 
Then, from the relations
\beqa
\tilde{\Delta } &=& \tilde{\Delta }_{1} + \tilde{\Delta }_{2} \ ,
\nonumber \\
\tilde{\Delta }_{1} &=& 
\frac{m^2}{p^2} \biggl[ \frac{\zeta}{\alpha ( 1-\alpha) } + 1 - \zeta  \biggr ]
> \frac{m^2}{p^2}\frac{\zeta }{\alpha } \ , \nonumber \\
\tilde{\Delta }_{2} &=&  
\frac{1}{\alpha} \biggl [\zeta\, \beta (1-\beta) 
+ \alpha (1-\zeta) \Bigl( \zeta(\beta - \eta)^2
+ \eta (1- \eta) \Bigr) \biggr] > \frac{\zeta }{\alpha } \beta (1-\beta ) \ ,
\nonumber
\eeqa
we can put a lower bound on $\tilde{\Delta }$ as
\beq
\tilde{\Delta } >
\frac{\zeta }{\alpha }\Bigl( \frac{m^2}{p^2} + \beta (1-\beta ) \Bigr) 
\ . \label{upper-del}
\eeq
Similarly, we decompose $F$ as $F = F_{1} + F_{2} + F_{3}$, where
\beqa
F_{1} &=& \frac{m^2}{p^2}\Bigl(1-\alpha(1-\alpha) \Bigr) \ ,
\nonumber \\ 
F_{2} &=& (1-\alpha) \beta(1-\beta)  \ ,
\nonumber \\
F_{3} &=& \alpha (1-\alpha )
\Bigl( (1-2\zeta)(\beta - \eta)^2 - \eta(1-\eta) \Bigr ) \  .
\nonumber 
\eeqa
Then, because $|F_{1}|<1$, $|F_{2}|<1$, $|F_{3}|<2$, we obtain $|F|< 4$.
Thus we have
%can put an upper bound on $|\mathcal{D}|$ as
\beq
| \mathcal{D} | < 4 \int_{0}^{1} \! d \alpha \int_{0}^{1} \! d \beta
\int_{0}^{1} \! d \zeta \int_{0}^{1} \! d \eta\, \int_{0}^{\infty} \! dt
\, \frac{| \ln \zeta |}{\alpha t}
\exp \biggl[ -t - \frac{\zeta }{4 t \alpha }
\Bigl( \frac{m^2}{p^2} + \beta (1-\beta ) \Bigr) p^2( \theta p)^2 \biggr ] \ .  
\nonumber  
\eeq
Next we change the integration variable $\zeta$ to
$\lambda \equiv 
 \frac{\zeta }{4 t \alpha }
\Bigl( \frac{m^2}{p^2} + \beta (1-\beta ) \Bigr) p^2( \theta p)^2 $,
and send the upper end of the $\lambda $-integral to $\infty$. This
yields
\beqa
| \mathcal{D} | &<& \frac{16}{p^2 (\theta p)^2} \int_{0}^{1} 
d \beta \,\frac{1}{\frac{m^2}{p^2} + \beta (1-\beta )}\,
\int_{0}^{1} d \alpha 
\, \int_{0}^{\infty} dt \int_{0}^{\infty} d \lambda  \, e^{-t - \lambda }
\nonumber \\ 
&\cdot& \biggl[ \, | \ln t | + | \ln \alpha  | + | \ln \lambda | + 
\left| \ln \Bigr( \frac{m^2}{p^2} + \beta (1-\beta )\Bigl ) \right| 
 + \left| \ln \frac{p^2(\theta p)^2}{4}\right| 
 \,  \biggr] \ . 
\nonumber 
\eeqa
Finally, using the inequalities
\beqa
\left| \, \ln \Bigr( \frac{m^2}{p^2} + \beta (1-\beta )\Bigl )\right| 
&<& \ln \frac{p^2}{m^2} \ ,
\nonumber \\  
\int_{0}^{1} d \beta \,\frac{1}{\frac{m^2}{p^2} + \beta (1-\beta )} &=& 
\frac{1}{ \sqrt{\frac{1}{4} + \frac{m^2}{p^2}} }
\ln \frac{\Bigr(\sqrt{\frac{1}{4} + \frac{m^2}{p^2}} 
+ \frac{1}{2} \Bigl)^2}{\frac{m^2}{p^2}}
< \,2\Bigr( 1 + \ln \frac{p^2}{m^2}\Bigl) \ ,
\nonumber 
\\
\int_{0}^{\infty } dt \, | \ln t| \, e^{-t} &\equiv& c < \frac{3}{2} \ ,
\nonumber
\eeqa
we arrive at the upper bound (\ref{upper-bound}) on $ | \mathcal{D}| $.
%obtain an upper bound on $ | \mathcal{D}| $ as
%% \beqa
%% | \mathcal{D} | &<& \frac{32}{p^2 ( \theta p)^2}\Bigl( 1 + \ln\frac{p^2}{m^2} \Bigr)
%% \biggr[ c + 1 + c + \ln\frac{p^2}{m^2} + \frac{p^2(\theta p)^2}{4} \biggl]
%% \nonumber \\
%% &<& \frac{32}{p^2 ( \theta p)^2}\Bigl( 1 + \ln\frac{p^2}{m^2} \Bigr)
%% \Bigr( 4 + \ln\frac{p^2}{m^2} + \ln\frac{ p^2 ( \theta p)^2}{4} \Bigr) \ .
%% \nonumber
%% \eeqa
%% \beq
%% | \mathcal{D} | 
%% < \frac{32}{p^2 ( \theta p)^2}\Bigl( 1 + \ln\frac{p^2}{m^2} \Bigr)
%% \Bigr( 4 + \ln\frac{p^2}{m^2} + \ln\frac{ p^2 ( \theta p)^2}{4} \Bigr) \ .
%% \eeq

\section{Derivation of a Stringent Upper Bound on 
$|\Pi_{\rm NP2}(p^2)|$}

\label{sec:derive-ub2}
\setcounter{equation}{0}
\renewcommand{\theequation}{B$\cdot$\arabic{equation}}

In this appendix we obtain an
upper bound on $|\Pi_{\rm NP2}( p^2)|$
that is more stringent than Eq.\ (\ref{upper-bound2})
and actually vanishes in the $p^2 \rightarrow \infty$ limit.
%a more stringent upper bound on $|\Pi_{\rm NP2}( p^2)|$
%than (\ref{upper-bound2}).
%This will allow us to confirm more directly
%that the type 2 diagram vanishes in the same limit.
Let us consider the integrand in the last line of (\ref{Pi-NP2}).
In the denominator we omit the $t^2\zeta $ term, 
and in the exponent we omit $\tilde{\Delta}_{\rm NP}$
and the term 
$(1-\zeta) \Bigl( \zeta(\beta - \eta)^2 + \eta (1- \eta) \Bigr)$
in $\tilde{\Delta}$.
%% other terms except 
%% \beq
%% \frac{m^2}{p^2}\Bigl[ \frac{\zeta }{\alpha (1-\alpha )} + 1-\zeta \Bigr]
%% + \frac{\zeta }{\alpha}\beta (1-\beta ) \nonumber
%% \eeq
%% of $\tilde{\Delta }$ in the exponent.
%By performing the integration explicitly, 
Thus, we obtain the upper bound
%% \beqa
%% | \Pi _{NP2}(p^2)| &<&  
%% \frac{g^4}{(4\pi)^d }
%% \Bigl( \frac{4}{\theta ^2}\Bigr)^{\frac{d}{2}}\frac{1}{(p^2)^5}
%% \int_{0}^{1} \,d\beta \int_{0}^{1} \,
%% d\alpha \int_{0}^{1} \,d\zeta \,
%% \frac{4!\, \alpha^2 (1-\alpha)^3 \zeta ^2(1-\zeta )}
%% {\Bigl[ \frac{m^2}{p^2}
%% \bigl( \zeta + \alpha (1-\alpha ) \bigr) 
%% + \zeta (1-\alpha )\beta (1-\beta ) \Bigr]^5}
%% \nonumber \\
%% &=& \frac{g^4}{(4\pi )^{d}}
%% \Bigr( \frac{4}{\theta ^2}\Bigl)^{\frac{d}{2}}\frac{1}{(m^2)^5}\,
%% G\Bigr(\frac{m^2}{p^2} \Bigl) \ ,
%% \nonumber
%% \eeqa
%% where the function $G(x)$ is defined by
%% \beq
%% G(x) = \frac{\frac{1}{2}x^2}
%% {\Bigr( \frac{1}{4}+ x \Bigl)^{\frac{3}{2}}}
%% \Biggl(  
%% \ln \frac{\sqrt{\frac{1}{4} + x} + \frac{1}{2}}
%% {\sqrt{\frac{1}{4} + x} - \frac{1}{2} } + \frac{1}{x}
%% \sqrt{\frac{1}{4} + x } \Biggr) \ .
%% \nonumber
%% \eeq
\beq
| \Pi _{\rm NP2}(p^2)|
< \frac{g^4}{(4\pi )^{d}}
\Bigr( \frac{4}{\theta ^2}\Bigl)^{\frac{d}{2}}\frac{1}{(m^2)^5}\,
G\Bigr(\frac{m^2}{p^2} \Bigl) \ ,
\label{stringentUB}
\eeq
where the function $G(x)$ is defined by
\beqa
G(x) &=& x^5 \int_{0}^{1} \,d\beta \int_{0}^{1} \,
d\alpha \int_{0}^{1} \,d\zeta \,
\frac{24\, \alpha^2 (1-\alpha)^3 \zeta ^2(1-\zeta )}
{\Bigl[ x \bigl( \zeta + \alpha (1-\alpha )(1-\zeta) \bigr) 
+ \zeta (1-\alpha )\beta (1-\beta ) \Bigr]^5}
\nonumber \\
&=& \frac{\frac{1}{2}x^2}
{\Bigr( \frac{1}{4}+ x \Bigl)^{\frac{3}{2}}}
\Biggl(  
\ln \frac{\sqrt{\frac{1}{4} + x} + \frac{1}{2}}
{\sqrt{\frac{1}{4} + x} - \frac{1}{2} } + \frac{1}{x}
\sqrt{\frac{1}{4} + x } \Biggr) \ .
\eeqa
Then, because $\lim_{x\rightarrow 0}G(x)=0$, Eq.\ 
(\ref{stringentUB}) confirms explicitly
that $\Pi _{\rm NP2}(p^2)$ 
vanishes in the $p^2 \to \infty $ limit.

%\bigskip


\begin{thebibliography}{99}
%%%%%%%%%%%%%%%%%%%%%%%%%%%%%%%%%%%%%%%%%%%%%%%%%%%%%%%%%%%%%
% Some macros are available for the bibliography:
%  o for general use
%    \JL : general journals                 \andvol : Vol (Year) Page
%  o for individual journal 
%    \AJ   : Astrophys. J.           \NC         : Nuovo Cim.
%    \ANN  : Ann. of Phys.           \NPA, \NPB  : Nucl. Phys. [A,B]
%    \CMP  : Commun. Math. Phys.     \PLA, \PLB  : Phys. Lett. [A,B]
%    \IJMP : Int. J. Mod. Phys.      \PRA - \PRE : Phys. Rev. [A-E]     
%    \JHEP : J. High Energy Phys.    \PRL        : Phys. Rev. Lett.
%    \JMP  : J. Math. Phys.          \PRP        : Phys. Rep.
%    \JP   : J. of Phys.             \PTP        : Prog. Theor. Phys.     
%    \JPSJ : J. Phys. Soc. Jpn.      \PTPS       : Prog. Theor. Phys. Suppl.
% Usage:
%  \PRD{45,1990,345}          ==> Phys.~Rev.\ \textbf{D45} (1990), 345
%  \JL{Nature,418,2002,123}   ==> Nature \textbf{418} (2002), 123
%  \andvol{B123,1995,1020}    ==> \textbf{B123} (1995), 1020
%%%%%%%%%%%%%%%%%%%%%%%%%%%%%%%%%%%%%%%%%%%%%%%%%%%%%%%%%%%%%

\bibitem{Sny} H.~S.\ Snyder,
% {\it Quantized space-time}, 
Phys.~Rev.\ \textbf{71} (1947), 38.\\
%\bibitem{Connes}
A.\ Connes, 
\emph{Noncommutative geometry} (Academic Press, 1990).

\bibitem{gravity} S.\ Doplicher, K.\ Fredenhagen and J.~E.\ Roberts,
%{\it The quantum structure of spacetime at the Planck scale 
%and quantum fields}, 
\CMP{172,1995,187}; hep-th/0303037.

\bibitem{rf:SW}
N.~Seiberg and E.~Witten, 
%``String theory and non-commutative geometry,''
\JHEP{09,1999,032};
hep-th/9908142.
%\hepth{9908142}

\bibitem{rf:MRS}
S.~Minwalla, M.~Van Raamsdonk and N.~Seiberg,
%``Noncommutative perturbative dynamics,"
\JHEP{02,2000,020};
hep-th/9912072.

\bibitem{GuSo} 
S.~S.~Gubser and S.~L.~Sondhi, 
%{\it Phase structure 
%of non-commutative scalar field theories}, 
\NPB{605,2001,395}; hep-th/0006119.

%%\cite{Bietenholz:2002vj}
\bibitem{Bietenholz:2002vj}
W.~Bietenholz, F.~Hofheinz and J.~Nishimura,
%``Simulating non-commutative field theory,''
Nucl.\ Phys.\ B (Proc.\ Suppl.\ ) {\bf 119} (2003), 941; hep-lat/0209021;
%%CITATION = HEP-LAT 0209021;%%
%\cite{Bietenholz:2002ev}
%\bibitem{Bietenholz:2002ev}
%W.~Bietenholz, F.~Hofheinz and J.~Nishimura,
%``Non-commutative field theories beyond perturbation theory,''
Fortsch.\ Phys.\  {\bf 51} (2003), 745; hep-th/0212258.
%%CITATION = HEP-TH 0212258;%%
%
\bibitem{AC}
%\bibitem{Ambjorn:2002nj}
J.~Ambj\o rn and S.~Catterall,
%\emph{Stripes from (noncommutative) stars,}
\PLB{549,2002,253}; hep-lat/0209106. \\
%%CITATION = HEP-LAT 0209106;%%
%\cite{Martin:2004un}
%\bibitem{Martin:2004un}
X.~Martin,
%``A matrix phase for the phi**4 scalar field on the fuzzy sphere,''
\JHEP{04,2004,077}; hep-th/0402230.
%%CITATION = HEP-TH 0402230;%%


%
%\cite{Bietenholz:2004xs}
\bibitem{Bietenholz:2004xs}
W.~Bietenholz, F.~Hofheinz and J.~Nishimura,
%\emph{Phase diagram and dispersion relation of the non-commutative 
%$\lambda \phi^4$ model in $d = 3$,}
\JHEP{06,2004,042}; hep-th/0404020.
%%CITATION = HEP-TH 0404020;%%



\bibitem{NCgauge} 
%\bibitem{rf:MVR}
M.~Van Raamsdonk,
%``The Meaing of Infrared Singulatities in Noncommutative Gauge Theories,"
\JHEP{11,2001,006}; hep-th/0110093. \\
%
%\bibitem{rf:AL}
A.~Armoni and E.~Lopez,
%``UV/IR Mixing via Closed Strings and Tachyonic Instabilities,"
\NPB{632,2002,240}; hep-th/0110113.



%\cite{Bietenholz:2004as}
\bibitem{Bietenholz:2004as}
W.~Bietenholz, F.~Hofheinz and J.~Nishimura,
%\emph{On the relation between 
%non-commutative field theories at $\theta = \infty$ 
%and large N matrix field theories,}
\JHEP{05,2004,047}; hep-th/0404179.
%ibid. \andvol{0405,2004,047}, hep-th/0404179.
%%CITATION = HEP-TH 0404179;%%


\bibitem{AMNS}
%\cite{Aoki:1999vr}
%\bibitem{Aoki:1999vr}
H.~Aoki, N.~Ishibashi, S.~Iso, H.~Kawai, Y.~Kitazawa and T.~Tada,
%``Noncommutative Yang-Mills in IIB matrix model,''
\NPB{565,2000,176}; hep-th/9908141. \\
%Nucl.\ Phys.\ B {\bf 565} (2000) 176
%[arXiv:hep-th/9908141].
%%CITATION = HEP-TH 9908141;%%
%
J.~Ambj\o rn, Y.~M.~Makeenko, J.~Nishimura and R.~J.~Szabo,
%\emph{Finite N matrix models of noncommutative gauge theory,}
\JHEP{11,1999,029}; hep-th/9911041;
%%CITATION = HEP-TH 9911041;%%
%\cite{Ambjorn:2000nb}
%\bibitem{Ambjorn:2000nb}
%\emph{Nonperturbative dynamics of noncommutative gauge theory,}
\PLB{480,2000,399}, hep-th/0002158;
%%CITATION = HEP-TH 0002158;%%
%\cite{Ambjorn:2000cs}
%\bibitem{Ambjorn:2000cs}
%\emph{Lattice gauge fields and discrete noncommutative Yang-Mills theory,}
\JHEP{05,2000,023}; hep-th/0004147.
%%CITATION = HEP-TH 0004147;%%



%\cite{Eguchi:1982nm}
\bibitem{Eguchi:1982nm}
T.~Eguchi and H.~Kawai,
%\emph{Reduction Of Dynamical Degrees Of Freedom In The Large N Gauge Theory,}
\PRL{48,1982,1063}.


%\cite{Gonzalez-Arroyo:1982hz}
\bibitem{Gonzalez-Arroyo:1982hz}
A.~Gonzalez-Arroyo and M.~Okawa,
%``The Twisted Eguchi-Kawai Model: A Reduced Model For Large N Lattice Gauge
%Theory,''
\PRD{27,1983,2397}.
%%CITATION = ,D27,2397;%%


%\cite{Ishibashi:1999hs}
\bibitem{Ishibashi:1999hs}
N.~Ishibashi, S.~Iso, H.~Kawai and Y.~Kitazawa,
%``Wilson loops in noncommutative Yang-Mills,''
\NPB{573,2000,573}; hep-th/9910004.
%%CITATION = HEP-TH 9910004;%%

\bibitem{rf:BHN}
 W.~Bietenholz, F.~Hofheinz and J.~Nishimura,
%``A non-perturbative study of gauge theory on a non-commutative plane,"
\JHEP{09,2002,009}; hep-th/0203151. 


\bibitem{rf:phi4}
I.~Ya.~Aref'eva, D.~M.~Belov and A.~S.~Koshelev,
%``Two-Loop Diagrams in Noncommutative $\varphi_{4}^{4}$ theory,"
\PLB{476,2002,431}; hep-th/9912075. \\
A.~Micu and M.~M.~Sheikh-Jabbari,
%``Noncommutative $\phi^4$ at Two Loops,"
\JHEP{01,2001,025}; hep-th/0008057. \\
W.~Huang,
%``Two-loop effective potential in noncommutative scaler field theory," 
\PLB{496,2000,206}; hep-th/0009067.


\bibitem{rf:phi3}
Y.~Kiem and S.~Lee,
%``Noncommtative Field Theory from String Theory: Two-loop Analysis,"
\NPB{594,2001,169}; hep-th/0008002. \\
Y.~Kiem, S.~Kim, S.~Rey and H.~Sato,
%``Anatomy of Two-Loop Effective Action in Noncommutative Field Theories,"
\NPB{641,2002,256}; hep-th/0110066.

%% \bibitem{rf:TM}
%%    T. Muta,``Foundation of Quantum Chromodynamics," 
%% World Scientific Lecture Notes in Physics - Vol.5. 

%% \bibitem{rf:tHV}
%%    G. 't Hooft and M. Veltman,
%% ``Regularization and Renormalization of Gauge Fields,"
%%    \NPB{44,1972,189}.

%\cite{Chepelev:1999tt}
\bibitem{Chepelev:1999tt}
I.~Chepelev and R.~Roiban,
%``Renormalization of quantum field theories on noncommutative R**d.  I:
%Scalars,''
\JHEP{05,2000,037}; hep-th/9911098.
%{\bf 0005} (2000) 037
%[arXiv:hep-th/9911098].
%%CITATION = HEP-TH 9911098;%%



\end{thebibliography}
\end{document}